\documentclass{llncs}
\usepackage{makeidx}  % allows for indexgeneration
\usepackage{graphics,graphicx}
\usepackage{amsmath,amsfonts,amssymb}
\usepackage{algorithm}
\usepackage{algpseudocode}
\newcommand{\xb}{\mathbf{x}}

\newcommand{\vb}{\mathbf{v}}
\newcommand{\Fb}{\mathbf{F}}
\newcommand{\Pb}{\mathbf{P}}
\newcommand{\Qb}{\mathbf{Q}}
\newcommand{\Ib}{\mathbf{I}}
\newcommand{\Ab}{\mathbf{A}}
\newcommand{\Ub}{\mathbf{U}}
\newcommand{\muB}{\mbox{\boldmath$\mu$}}
\begin{document}

\title{Robust TMA using the possibility particle filter}
\titlerunning{TMA using the possibility particle filter}  % abbreviated title (for running head)
\author{Branko Ristic\inst{1}%\orcidID{0000-1111-2222-3333}
\and
Jeremie Houssineau\inst{2}
%\orcidID{1111-2222-3333-4444}
\and
Sanjeev Arulampalam\inst{3}
%\orcidID{2222-3333-4444-5555}
%\and
%A. Bishop\inst{4}
%\orcidID{3333-4444-5555-6666}
}
\authorrunning{B. Ristic et al.} % abbreviated author list (for running head)
\tocauthor{B. Ristic, J. Houssineau, S. Arulampalam}
\institute{RMIT University, Australia, \\ \email{branko.ristic@rmit.edu.au}
\and
National University of Singapore, Singapore, \\ \email{houssineau.j@gmail.com}
\and
Defence Science and Technology, Australia, \\ \email{sanjeev.arulampalam@dst.defence.gov.au}
}

\maketitle              % typeset the title of the contribution

\begin{abstract}
The problem is target motion analysis (TMA), where the objective is to estimate the state of a moving target from noise corrupted bearings-only measurements. The focus is on recursive TMA, traditionally solved using the Bayesian filters (e.g. the extended or unscented Kalman filters, particle filters). The TMA is a difficult problem and may cause the algorithms to diverge, especially when the measurement noise model is imperfect or mismatched. As a robust alternative to the Bayesian filters for TMA, we propose the recently introduced possibility filter. This filter is implemented in the sequential Monte Carlo framework, and referred to as the possibility particle filter. The paper demonstrates its superior performance against the standard particle filter in the presence of a model mismatch, and equal performance in the case of the exact model match.
\end{abstract}
\section{Introduction}
Target Motion Analysis (TMA) algorithms estimate the state of a moving target, such as its position and velocity, from noise corrupted measurements of bearings to the target, provided by an acoustic
sensor. TMA plays an important role in submarine combat systems and has a long history of developments \cite{lingren1978position,nardone1984fundamental}. The modern emphasis is on recursive Bayesian methods (such as the  extended or unscented Kalman filter, or the particle filter \cite{pfbook}),  because they estimate the  entire posterior probability density function of the state and thereby provide a measure of uncertainty to derived point estimates.  TMA is a difficult nonlinear filtering problem, because the target range remains unobservable until the observer performs an appropriate manoeuver \cite{fogel_gavish_88}. Consequently,  the recursive TMA algorithms can occasionally fail and cause track divergence. This is particularly true in situations where the nonlinearity is high or the measurement noise model is imperfect or mismatched.

The possibility filter has recently been introduced  as a special instance of a class of outer measure based stochastic filters and smoothers \cite{houssineau_bishop_17}. Bayesian filtering style analytic expressions for prediction and update of outer measures are available and their numerical implementation is discussed in \cite{houssineau_bishop_17,houssineau2017sequential,bishop2017spatial}. The motivation for using outer measures instead of the probabilistic framework is to provide a more generalised representation of uncertainty, capable of handling in a rigorous mathematical manner the situations of ignorance or partial knowledge. The possibility filter is adopted in this paper for TMA  in the hope that it will provide robustness against nonlinearities and the mismatch in the measurement model. Because there is no closed form analytic solution to  the prediction and update equations of the possibility filter for TMA,  an approximate solution based on the sequential Monte Carlo estimation framework is implemented, following \cite{houssineau2017sequential}.

%The paper is organised as follows. Sec. \ref{II} describes the TMA problem. Sec. \ref{III} presents a short overview of the possibility filter, including the filtering equations. Sec. \ref{IV} explains the Monte Carlo implementation of the possibility particle filter (PPF), followed by numerical results in Sec. \ref{V}. Finally, a summary of the paper is given in Sec. \ref{VI}.

%
\section{TMA modeling and formulation}
\label{II}
Our goal is to estimate the state of a moving object. The state at time $t_k$ ($k=0,1,2,\dots$) is fully characterised by the state vector $\xb^t_k\in\mathbb{X}_k\subseteq \mathbb{R}^{d}$ (where $\mathbb{X}_k$ is the state space). We adopt the two-dimensional Cartesian coordinate system and define the state vector as
\begin{equation}
\xb^t_k = \left[\begin{matrix}x^t_k & \dot{x}^t_k & y^t_k &
\dot{y}^t_k\end{matrix}\right]^\intercal.
\end{equation}
where $(x^t_k,y^t_k)$ and $(\dot{x}^t_k,\dot{y}^t_k)$ represent the target position and velocity, respectively.
The observer (ownship) is also moving in the same coordinate system. Its state vector is assumed known and correspondingly defined as $\xb^o_k = \left[\begin{matrix}x^o_k & \dot{x}^o_k & y^o_k &
\dot{y}^o_k\end{matrix}\right]^\intercal$.
The target motion model and the observation model will be expressed in terms of the {\em relative} state vector
$\xb_k := \xb^t_k - \xb^o_k =\left[\begin{matrix}x_k & \dot{x}_k & y_k &
\dot{y}_k\end{matrix}\right]^\intercal$.
The target motion model is adopted as the nearly constant velocity (CV) model \cite{barshalom_et_al_01}:
\begin{equation}
\xb_{k} = \Fb_{k-1}\xb_{k-1} -\Ub_{k,k-1}+ \vb_{k-1} \label{e:dyn_eq}
\end{equation}
where $\Fb_{k-1}$ is the transition matrix, $\Ub_{k,k-1}$ is a known
deterministic matrix taking into account the effect of observer
accelerations. Dynamic (or process) noise in (\ref{e:dyn_eq}) is captured by $\vb_{k-1}$, which  represents an uncertain variable (see \cite{houssineau_bishop_17} for details), introduced to model the uncertainty due to randomness and/or incomplete probabilistic description of the target motion model. In the next section we discuss the characterisation of $\vb_{k-1}$ with a possibility distribution.
 Matrices $\Fb_k$ and $\Ub_k$ are defined as:
\begin{equation}
\Fb_{k-1} =  \Ib_2 \otimes \left[\begin{matrix}1 & T_{k-1} \\ 0 &
1\end{matrix}\right], \hspace{5mm}\Ub_{k+1,k}  =  \xb_{k+1}^o-\Fb_k\xb_{k}^o,
\end{equation}
where $\otimes$ is the Kroneker product, while $T_{k-1}=t_{k}-t_{k-1}$ is the
sampling interval. We refer to $k$ as the
discrete-time index or the scan. By adopting a constant sampling interval $T_{k-1}=T=const$, notation
simplifies to $\Fb_{k-1}=\Fb$.

Bearing measurement is related to the target state at time $t_k$  as follows:
 \begin{equation}
 z_k = h(\xb_k)+w_k
 \label{e:meas_eq}
 \end{equation}
 where $w_k$ is an uncertain variable modelling the measurement noise process, independent \cite{houssineau_bishop_17} from uncertain variable $\vb_k$. Characterisation of $w_k$ using a possibility distribution will be discussed in the next section. The nonlinear function $h(\cdot)$ in (\ref{e:meas_eq}) is the four-quadrant inverse tangent function:
 \begin{equation}
 h(\xb_k) = \mbox{atan2}(x_k,y_k), \label{e:arctan2}
 \end{equation}
resulting in the true target bearing  at time $t_k$.
 %The likelihood function can then be expressed as $\ell(z_k|\xb_k)=\mathcal{N}(z_k;h(\xb_k),\sigma^2)$.

\section{Possibility filter}
\label{III}

Consider an uncertain variable $\xb$ on the state space $\mathbb{X}$.
 The concept of the possibility distribution was introduced in the seminal paper \cite{zadeh_78}. Let $\Ab$ be a (nonfuzzy) subset of $\mathbb{X}$ and let $\Pi$ be a possibility distribution associated with an uncertain variable $\xb\in\mathbb{X}$. Then the possibility measure of $\Ab$ is defined as a number in the interval $[0, 1]$ given by:
$Poss(\xb\in \Ab) = \sup_{\xb\in \Ab} \pi(\xb)$,
where $\pi(\xb)$ is the possibility distribution function (pdf) of $\Pi$.  The pdf $\pi:\mathbb{X}\rightarrow [0,1]$ is a membership function determining the fuzzy restriction on $\xb$. Any probability density function $p(\xb)$ can be turned into a possibility distribution function $\pi(\xb)$ as follows: $\pi(\xb) = p(\xb) / \sup_{\xb\in \mathbb{X}} p(\xb)$. We will restrict in this paper on a Gaussian pdf defined as:
\begin{equation}
\pi(\xb) = \bar{\mathcal{N}}(\xb;\muB,\Pb) = \exp\left(-\frac{1}{2}(\xb-\muB)^\intercal\Pb^{-1}(\xb-\muB)\right)
\label{e:Gauss}
\end{equation}
for some $\muB\in\mathbb{R}^d$ and for some $d\times d$ positive definite matrix $\Pb$ with real coefficients. With abuse of language, we will refer to $\muB$ and $\Pb$ as to the mean\footnote{The possibilistic mean value has been defined as a closed interval \cite{dubois1987mean}, although other interpretations exist.}
%,carlsson2009possibilistic}.
 and variance of pdf $ \bar{\mathcal{N}}(\xb;\muB,\Pb)$.

 Referring to the TMA problem, the state vector $\xb_k$ is treated as an uncertain variable whose uncertainty is represented by the pdf over the state space $\mathbb{X}_k$. Let us introduce the concept of a posterior pdf $\pi(\xb_k|z_{1:k}$), that is the conditional pdf of $\xb_k$ given all measurements $z_{1:k}\equiv z_1,z_2,\dots,z_k$ up to time $t_k$.
The problem of recursive TMA is to estimate sequentially the posterior pdf, assuming that the initial pdf (at time $t_1$) $\pi(\xb_1)$, is known.

The prediction and update equations of the possibility filter are presented next. Suppose the posterior pdf at time $k-1$, that is $\pi(\xb_{k-1}|z_{1:k-1})$, is available. The prediction equation explains how to compute the pdf at time $k$ based on the target dynamic model only. It is given by  \cite{houssineau_bishop_17}:
\begin{equation}
\pi(\xb_k|z_{1:k-1}) = \sup\limits_{\xb_{k-1}\in\mathbb{X}_{k-1}} \varphi(\xb_k|\xb_{k-1})\,\pi(\xb_{k-1}|z_{1:k-1})
\label{e:pred_eq_pf}
\end{equation}
where $\varphi(\xb_k|\xb_{k-1})$ is the conditional pdf describing the transition from $\xb_{k-1}$ to $\xb_k$.  This transition was described by (\ref{e:dyn_eq}) for the TMA problem. Let us model process noise $\vb_k$ in (\ref{e:dyn_eq}) as an uncertain variable characterised by the Gaussian pdf  (\ref{e:Gauss}), with zero-mean and covariance matrix
\begin{equation}
\Qb = \Ib_2 \otimes q
\left[\begin{matrix} \frac{T^3}{3} &
\frac{T^2}{2}\\\frac{T^2}{2} & T\end{matrix}\right],
\end{equation}
where $q$ determines the intensity of process noise. The conditional pdf $\varphi(\xb_k|\xb_{k-1})$ in this case can be expressed by a Gaussian pdf (\ref{e:Gauss}) with mean $\Fb\xb_{k-1}-\Ub_{k,k-1}$ and covariance $\Qb$. Note that the prediction equation (\ref{e:pred_eq_pf})  is the analogue
of the Chapman-Kolmogorov equation in the standard
Bayesian filtering \cite{jazwinski_70}, except that: (i) the integral is replaced by the
supremum and (ii) the probability density functions are replaced by the
possibility distribution functions.

The update step of the possibility filter ``corrects''  the prediction  $\pi(\xb_k|z_{1:k-1})$ using the information contained in the new measurement $z_k$. The update equation is given by  \cite{houssineau_bishop_17}:
\begin{equation}
\pi(\xb_k|z_{1:k}) = \frac{g(\xb_k,z_k)\,\pi(\xb_k|z_{1:k-1})}{\sup_{\xb\in\mathbb{X}_k} \left[g(\xb,z_k)\,\pi(\xb|z_{1:k-1})\right]}
\label{e:upd_eq_pf}
\end{equation}
Equation (\ref{e:upd_eq_pf}) is the analogue of the Bayes' theorem, with the exception that: (i) the supremum replaces the integral, and (ii) the probability density functions are replaced by the
possibility distribution functions. The term $g(\xb_k,z_k)$ in (\ref{e:upd_eq_pf}) represents the likelihood function. Assuming the uncertain variable  $w_k$ in (\ref{e:meas_eq})  is characterised by a Gaussian pdf  (\ref{e:Gauss}), with zero-mean and variance $\sigma^2$, the likelihood function  $g(\xb_k,z_k)$ is also a Gaussian pdf  (\ref{e:Gauss}), with the mean value  $h(\xb_k)$ and variance $\sigma^2$.

A point estimate of the state vector at time $t_k$ can be computed from the pdf  $\pi(\xb_k|z_{1:k-1})$ as the maximum a posteriori (MAP) estimate, i.e.
\begin{equation}
\widehat{\xb}_{k|k} = \arg\max\limits_{\xb_k\in\mathbb{X}_k}  \pi(\xb_k|z_{1:k}).
\end{equation}

There is no analytic closed form solution to TMA using the possibility filter -  hence we must resort to numerical approximations. One option would be a grid-based method \cite{bishop2017spatial}: the state space would be divided into a regular $d$-dimensional grid and the value of $\pi(\xb_k|z_{1:k})$ would recursively be computed using the prediction and update equations in each node of this grid, as the time progresses.  It is well known, however, that the grid-based methods suffer from the curse of dimensionality. Instead we propose to implement the possibility filter using the sequential Monte Carlo method (SMC).   SMC algorithms have become widespread in Bayesian estimation and their properties, extensions and applications have been studied in numerous publications \cite{smcbook,pf_tute,pfbook,cape_07}
 %doucet2009tutorial}.
% The key feature of the SMC method is to propagate in time the samples approximating the posterior probability distribution.

\section{Monte Carlo approximation of the possibility filter}
\label{IV}

The practical implementation of the possibility filter for stochastic filtering is based on adaptation of the SMC method for propagation of the support points of the possibility distribution function over time. These points are known as {\em particles} and hence the resulting filter is referred to as the possibility particle filter (PF).

The problem we face is that sampling does not apply directly to possibility distributions. Instead, for a given possibility distribution function $\pi$, samples must be drawn from a probability density function $p$ which is induced by $\pi$. While there is an infinite number of ways one can construct $p$ from $\pi$, the natural solution is the one that results in the least informative $p$. This can be achieved by application of the  {\em maximum entropy} principle, formally stated as follows:  given a possibility distribution function $\pi$, we are after the probability density function $p^*$ that will maximise the differential entropy (defined as $H(p) = \mathbb{E}[-\ln p(\xb)]$), subject to constraints that: (i) $p^*$ integrates to $1$ and (ii) $p^*$ is lower or equal to $\pi$ for every $\xb\in\mathbb{X}$.

\begin{figure}[tbh]
\centerline{\includegraphics[height=3.2cm]{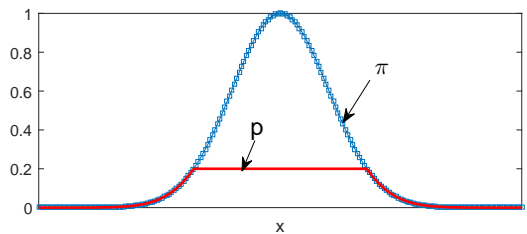}}
\caption{Water-pouring operation for a continuous distribution in one dimension} \label{f:1}
\end{figure}

The described constrained optimisation is numerically difficult to solve and instead we will adopt its approximation, achieved by what we refer to as the ``water pouring operation''. First note that if we were to ignore condition (ii) above, the uniform probability distribution would be the solution. By the water-pouring operation, we construct $p^*$ as the density closest to the unform, with equal support as $\pi$, which integrates to $1$. This operation will be denoted $\mathcal{P}^*(\pi)$; it is illustrated in Fig. \ref{f:1} for one dimensional continuous uncertain variable. The same principle can be applied to a discrete-valued uncertain variable.

A sample $\xb^j$ drawn from the probability distribution $\mathcal{P}^*(\pi)$ has to be weighted by $w^j=\pi(\xb^j)$. The set of $N\gg 1$ weighted samples $ \{(w^j,\xb^j)\}_{1\leq j\leq N}$ is then a Monte Carlo approximation of the pdf $\pi(\xb)$. Note that the density of samples has no impact on the supremum, as opposed to the integral in the standard formulation.

Pseudo-code of the possibility PF is given in Alg. \ref{a:1}.  Lines 1 and 2 execute the initialisation stage of the possibility PF. The loop between lines 3-18 is carried out at each discrete-time step. While the code in Alg. \ref{a:1} should be self-explanatory, except for the line 13; here we apply the water-pouring operation to find the probability mass function which corresponds to the weighted particle set $\{(\widehat{w}^j_k,\xb^j_{k-})\}_{1\leq j\leq N}$. The loop in lines 12-16 performs resampling in order to focus the computational effort of the possibility PF on the areas of the state space with non-negligible likelihood.

\begin{algorithm}[htb]
\caption{Pseude-code of the possibility particle filter}
{
\begin{algorithmic}[1]
\State  $\xb_{1}^j \sim \mathcal{P}^*(\pi_1)$, $\tilde{w}_{1}^j = \pi_1(\xb_{1}^j)$, for
$j=1,\cdots,N$ \Comment Initialisation
\State  $w_{1}^j =  \tilde{w}_{1}^j \; / \max\limits_{1\leq i \leq N}  \tilde{w}_{1}^i$, for $j=1,\cdots,N$ \Comment Normalisation of weights
\For {$k=2,3,\cdots$}
    \For {$j=1,2,\cdots,N$}  %\Comment Prediction
        \State  $\xb_{k-}^j \sim \mathcal{P}^*\left(\varphi(\cdot|\xb_{k-1}^j)\right)$
        \State $w_{k-}^j = \varphi(\xb_{k-}^j|\xb^j_{k-1}) w_{k-1}^j$
    %\EndFor
    %\State  $w_{k-}^j =  \tilde{w}_{k-}^j \; / \max\limits_{1\leq i \leq N} \tilde{w}_{k-}^i$, for $j=1,\cdots,N$
    %\For {$j=1,2,\cdots,N$}  \Comment Update
        \State $\tilde{w}^j_{k} = w^j_{k-}\,g(\xb^j_{k-},z_k)$
    \EndFor
    \State $j_{\max} = \arg\max_{1\leq j \leq N}  \tilde{w}_{k}^j$
    \State Output the MAP estimate: $\widehat{x}_{k|k} \equiv \xb^{j_{\max}}_{k-}$
    \State  $\widehat{w}_{k}^j =  \tilde{w}_{k}^j \; / \tilde{w}_k^{j_{\max}}$, for $j=1,\cdots,N$
    \For {$j=1,2,\cdots,N$}  \Comment Resampling
        \State $a_j \sim \mathcal{P}^*(\{\widehat{w}_{k}^j\}_{1\leq j \leq N})$
        \State $\xb^j_{k} = \xb^{a_j}_{k-}$
        \State $\tilde{w}_k^j = \widehat{w}_{k}^{a_j}$
    \EndFor
   \State  $w_{k}^j =  \tilde{w}_{k}^j \; / \max\limits_{1\leq i \leq N}  \tilde{w}_{k}^i$, for $j=1,\cdots,N$
\EndFor
\end{algorithmic}
} \label{a:1}
\end{algorithm}

\section{Numerical results}
\label{V}
\begin{figure}[htb]
\centerline{\includegraphics[height=4cm]{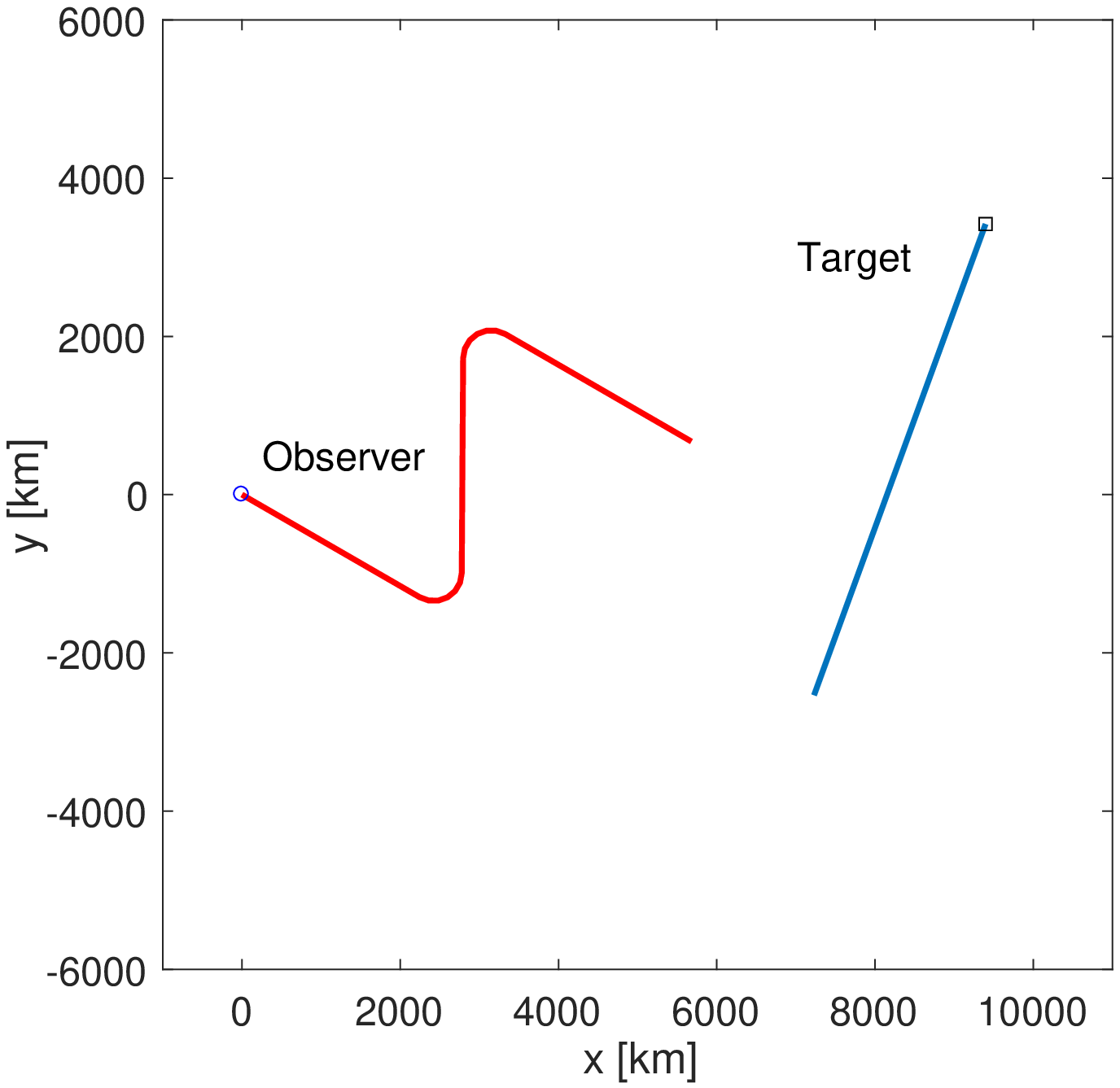}\includegraphics[height=4cm]{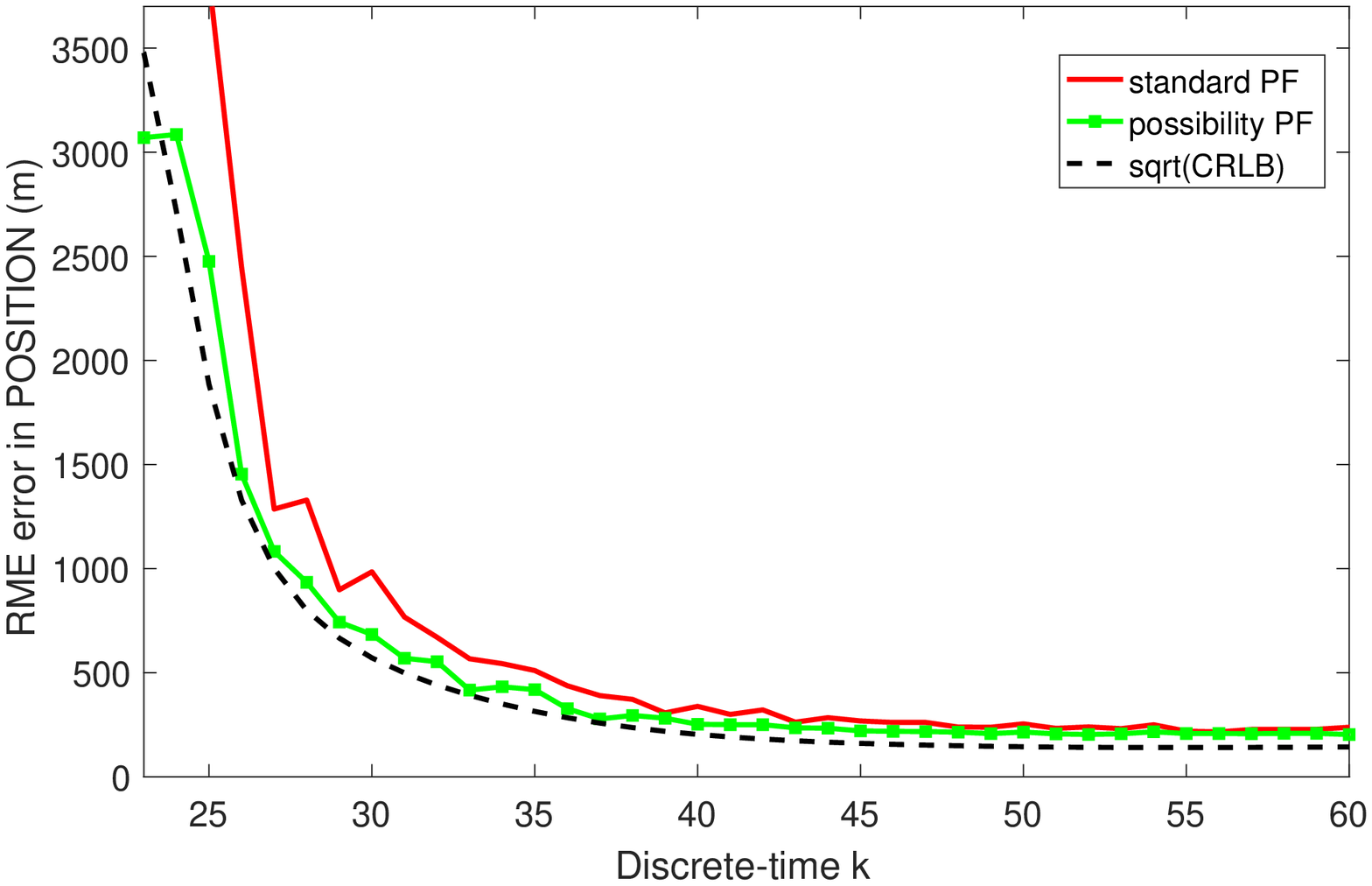}}
 \caption{(a) Top-down view of the TMA scenario; (b) RMS positional errors obtained by averaging over 500 Monte Carlo runs.  } \label{f:2}
\end{figure}

The scenario used in simulations is inspired by sonar underwater surveillance and plotted in Fig. \ref{f:2}.(a). The sampling interval was $T=40$ s. The initial pdf $\pi(\xb_1)$ in the possibility PF is constructed using measurement $z_1$ as a Gaussian pdf (\ref{e:Gauss}), with the mean $\bar{\xb} = [\bar{R}\sin z_1\;\;-\dot{x}_1^o \;\; \bar{R}\cos z_1\;\; -\dot{y}_1^o]^\intercal$ and covariance
\begin{equation}
\mathbf{P}_1 = \left[\begin{matrix} \sigma_x^2 &  0 & \sigma_{xy} & 0\\
                                          0 & \sigma_{\dot{x}}^2 & 0 & 0\\
                                          \sigma_{xy}  & 0    & \sigma_y^2 & 0\\
                                          0  & 0     & 0 &\sigma_{\dot{y}}^2\end{matrix}\right]
                                          \label{e:P0x}
                                          \end{equation}
where \cite{pfbook}:
$\sigma_{x}^2  =  \sigma_R^2\;(\cos z_1)^2 + \bar{R}^2\;\sigma^2\;(\sin z_1)^2$,
$\sigma_{y}^2  =  \sigma_R^2\;(\sin z_1)^2 + \bar{R}^2\;\sigma^2 \;(\cos z_1)^2$, and
$\sigma_{xy}  =  (\sigma_R^2 - \bar{R}^2 \;\sigma^2) \;\sin z_1\;\cos z_1$.
The values used in the possibility PF are: $\bar{R}=10$km, $\sigma_R = 3.5$km, $\sigma= 1^o$.

The first set of Monte Carlo simulation runs was carried out by drawing the measurement noise samples $w_k$ in (\ref{e:meas_eq}) from the zero-mean Gaussian probability density function, with the standard deviation $\sigma=1^o$. In this experiment we compare the standard particle filter (PF) for TMA \cite{pfbook} with the possibility PF. The standard PF assumes the zero-mean Gaussian measurement noise model with the value of standard deviation $\sigma=1^o$ (i.e. exactly matched model to the way we generate the bearing measurements). The possibility PF, on the other hand, models the measurement noise with the zero-mean Gaussian possibility distribution (\ref{e:Gauss}), with $\sigma=1^o$. The RMS positional errors, obtained from 500 independent Monte Carlo runs,  are shown in Fig. \ref{f:2}.(b), against the theoretical Cramer-Rao lower bound (CRLB) \cite{pfbook}. The number of particles in both the standard PF and the possibility PF was set to $N=20000$. None of the two contesting filters diverged in 500 runs.

The objective of the first set of Monte Carlo runs is to confirm that in the case where the measurement models are matched, the possibility PF is not inferior to the standard PF. Indeed we see from Fig. \ref{f:2}.(b) that, when we use a large number of particles, their RMS errors are matched and very close to the best achievable error performance indicated by the CRLB.

The second set of Monte Carlo runs considers the mismatched measurement noise models and a smaller number of particles. We continue to use the same two particle filters (the standard PF and the possibility PF), but now the noise samples $w_k$  in (\ref{e:meas_eq}) are actually drawn from a zero-mean Student-t distribution with $\sigma=1$ and degrees-of-freedom parameter $\nu$. The results obtained from 500 Monte Carlo runs are shown in Table \ref{t:1}. Due to the model mismatch and a smaller number of particles, the two contesting filters occasionally can diverge. A divergence is declared if the positional error at the end of the scenario is larger than 1 km. Table \ref{t:1}  presents the percentage of divergent runs for different values of $N$ and $\nu$. Note that as $\nu$ is increased, the tails of the Student-t distribution are reduced, and in the limit $\nu\rightarrow\infty$, it becomes equal to the Gaussian distribution. From Table \ref{t:1} one can observe that the possibility PF is more robust than the standard PF. The most dramatic improvement can be noted for $\nu\geq 5$, when the model mismatch is relatively mild. For example, at $\nu=8$, the percentage of divergent runs is reduced more than 8 times. The possibility PF is also more robust against the reduction of the number of particles $N$; for example, it never diverges even with only $N=2000$ at $\nu\rightarrow \infty$.

\begin{table}[tbh]
\caption{Percentage of divergent runs}
\centering
\begin{tabular}{ccccccccccc}
\hline
& \multicolumn{4}{c}{$N=2000$} &&& \multicolumn{4}{c}{$N=5000$}
\\\cline{2-5}\cline{8-11}
 $\;\;\;\nu\;\;\;$    & $\;3\;$ & $\;5\;$  & $\;8\;$ & $\;\infty\;$ &&& $\;3\;$ & $\;5\;$  & $\;8\;$ & $\;\infty\;$ \\
\hline
standard PF\hspace{2mm} & $\;45.2\;$  & $\;23.4\;$  & $\;16.4\;$   & $\;7.2\;$ &&& $\;33.4\;$ & $\;13.2\;$  & $\;7.0\;$  & $\;1.0\;$\\
possibility PF\hspace{2mm} & $25.0$ & $5.4$  & $2.0$  & $0.0$   &&& $19.0$ & $4.2$  &  $0.8$ &  $0.0$ \\
\hline
\end{tabular}
\label{t:1}
\end{table}

\section{Conclusions}
\label{VI}

The paper introduced the possibility particle filter as a robust alternative to the Bayesian filters for recursive target motion analysis. In comparison with the standard particle filter, the numerical results demonstrated: (i) an equal error performance in the case when the measurement noise models match; (ii) a significant reduction in divergences when the measurement models are mismatched or the number of particles is small.

\bibliographystyle{unsrt}
\bibliography{track}
%%
%% ---- Bibliography ----
%%
%\begin{thebibliography}{5}
%%
%\bibitem {clar:eke}
%Clarke, F., Ekeland, I.:
%Nonlinear oscillations and
%boundary-value problems for Hamiltonian systems.
%Arch. Rat. Mech. Anal. 78, 315--333 (1982)
%
%
%\end{thebibliography}

\end{document}